\documentstyle[12pt,aaspp4]{article}
\lefthead{Bouwens & Silk}
\righthead{Interpretative Evaluation of the Gronwall and Koo Model}
\begin{document}
\title{Passive Evolution: Are the Faint Blue Galaxy Counts Produced by a
 Population of Eternally Young Galaxies?}
\author{Rychard J. Bouwens}
\affil{Department of Physics, University of California, Berkeley, CA 94720;
       bouwens@astro.berkeley.edu}
\and
\author{Joseph Silk}
\affil{Departments of Astronomy and Physics, and Center for Particle
Astrophysics, University
  of California, Berkeley, CA 94720; silk@astro.berkeley.edu}

\begin{abstract}

A constant age population of blue galaxies, postulated in the model of 
Gronwall \& Koo (1995), seems to provide
 an  attractive explanation of the excess of very blue 
galaxies in the deep galaxy counts. Such a population may be  generated by a
set of galaxies with cycling star formation rates,  or at the other extreme,
be maintained by the continual formation of new galaxies which fade after they 
reach the age specified in the Gronwall and Koo model.  For both of these
hypotheses, we have calculated  the luminosity functions including the
respective selection criteria,
the redshift distributions,
 and the number counts in the $B_J$ and $K$ bands.
We find a substantial excess in the number of galaxies at low redshift
 ($0 < z < 0.05$) over that observed in the CFH redshift survey
(\cite{lil95}) and at
the faint end of the Las Campanas luminosity function
(\cite{lin96}).
Passive or mild evolution  fails to account
for the deep galaxy counts because of the implications for low redshift
 determinations of  the $I$-selected redshift distribution 
and the $r$-selected luminosity function  in  samples where the faded 
counterparts of the star-forming galaxies would be detectable.  

\end{abstract}

\section{Introduction}

That the deep galaxy counts require an extensive  blue population of faint
galaxies is undisputed.  A variety of different models invoke large scale
 merging or new galactic populations to explain this excess.  Others claim that
 this excess can be explained simply with mild (essentially passive)
 evolution and that the introduction of merging and new populations is
 unnecessary ({\it e.g.}  Gronwall \& Koo 1995, hereinafter GK; Pozzetti, Bruzual, \& Zamorani 1996).
In particular,  GK provide an  attractive explanation of  the deep galaxy counts
by determining the local luminosity functions (LFs) for galaxies in 11 different
morphological classes  so
as to also fit the galaxy redshift distributions and
broadband colors.  For
eight of the eleven morphological classes, GK allow the luminosities and colors
of the galaxies
to evolve in a way that is consistent with their construction of their
morphological classes.
GK require that the remaining three classes be very blue and completely
 non-evolving.  Maintaining this population of
 blue galaxies requires either the continual formation of
blue galaxies, in which case a remnant of fading galaxies would be left, or
a cycling
star formation rate in these galaxies, in which case the colors would not
seem to be
the same unless there was a significant amount of time between bursts
(\cite{bab95}).  In this paper, we impose the aforementioned physical
interpretations on the non-physical
blue galaxies in the GK model, and we assess their credibility.

\section{Results}

 In our physical interpretations of the GK model, we maintain GK's population
 of non-physical blue galaxies from $z = 0$ to $z = 1$.  We believe that $z=1$
 is early enough to account for a predominance of blue galaxies seen at
 faint magnitudes--which the GK model attempts to explain with
 a population of  non-physical blue galaxies--but not so early as to produce an
 unreasonably high number of faded galaxies. 
We interspersed periods of constant star formation of duration equal to that
 given by GK for the non-physical blue galaxies with periods of no star
 formation.  If the B-V colors specified for the non-physical blue galaxies in
 the GK model are not to be more than 0.10 magnitude different from the colors
 specified in the GK model during the active star formation phase, we find
 that
 the periods of no star formation have to be at least 1.2 Gyr for GK's bluest
 morphological type and 2.0 Gyr for GK's second bluest morphological type.

 We consider four models, which bear out a range of different physical
 interpretations of the GK models.  In Model A, the bluest class of galaxies 
 corresponding to class 1 of the GK model undergoes a period of constant
 star formation for 0.4 Gyr followed by a 1.2 Gyr period with no star
 formation, and    repeats this cycle indefinitely.
  The second bluest
 class of galaxies corresponding to class 2 of the GK model undergoes
 a period of constant star formation for 2 Gyr followed by a 2 Gyr period
 with no star formation and again repeats this cycle indefinitely.  Models B, C,
 and D are similar to model A except the bluest class of galaxies have 2.4 Gyr,
 4.8 Gyr, and an infinite period of time, respectively, separating the
 bursts of star
 formations, and the second bluest class of galaxies have 6 Gyr,
 12 Gyr, and an infinite period of time, respectively, separating the bursts of
 star formation.  To maintain this constant population of young blue galaxies
 in the GK model, we employed several sets of these galaxies with the star
 formation timed so that exactly one set of these galaxies would be
 undergoing their burst of constant star formation at any given  time.
 Obviously, in
 models where galaxies cycle more frequently, a much smaller set of galaxies
 is required and in models where galaxies cycle less frequently, a much larger
 set of galaxies is required to maintain this population of very blue galaxies.
 We assumed a Salpeter (1955) IMF
 with upper and lower mass limits of $0.1 M_{\sun}$
 and $125 M_{\sun}$, respectively, and aged the galaxies using a relatively
 current (1995) version of the spectral evolution code of Bruzual \& Charlot
 (1993).
  In accordance with the GK model,
  we assume a SMC extinction law  (\cite{bou85}) and $E(B-V) = 0.1$ for
 all but  the two reddest morphological types, and we assume that
 $H_{0} = 50 \ \rm{km\:s}^{-1} \rm{Mpc}^{-1}$ throughout.

 GK do not specify the surface brightness properties of their sample, and we
 adopt
 values derived from local observations and the relation between star formation
 history and morphological type.  We take the three reddest classes of
 objects in
 GK's model (GK classes 9-11; $B-V \geq 0.85$) to represent elliptical
 galaxies with a
 de Vaucouleurs profile with  intrinsic half-life radii determined by
 Bingelli, Sandage, \& Tarenghi (1984).
 We take the next three reddest classes (GK class 6-8; $0.65 \leq B-V \leq
0.85$)
 to represent Sa-Sc galaxies with surface brightnesses given by Freeman's law
 (\cite{fre70}) on which we superimpose an exponential
profile
 for the bulge with a total flux equal to one quarter that of the disk and a
scale
 length equal to 0.082 that of the disk (\cite{cou96}).
 We take the five bluest classes to be irregular/late spiral galaxies
 (GK class 1-5; $B-V \leq 0.65$) with a surface brightness for the disk
 identical to that for the Sa-Sc galaxies.
  Since GK do not specify a distinct star formation,
 for simplicity, 
we assume the same star formation history for the bulges and the disks.
 We mimic seeing and other smearing effects by convolving the angular
 profiles of
 each simulated galactic image with a Gaussian point spread function.
 We consider the disk galaxies to be oriented at an ensemble of
 incident angles, and we apply the selection criteria used in various
 determinations of the LF in a way very similar to that outlined in 
 Yoshii (1993).

 From these models, we calculated the luminosity functions that would have been
 determined by Loveday et al.\ (1992), Marzke, Huchra, \& Geller (1994),
 Lin et al.\ (1996), and Mobasher et al.\ (1993) from
 the APM, CfA, Las Campanas, and Anglo-Australian redshift surveys,
 respectively.  Our calculations are shown in Figure \ref{fig1}.
  Because the selection criteria for these surveys are often
 variable from field to field or even somewhat unclear, we consider the
selection criteria
 we have employed to be ``average'' estimates of the true selection criteria.
For the APM survey, we selected galaxies with apparent
 magnitudes between 15 and 17.15 $B_J$ mag and whose surface brightness is
at least
 24.5 $B_J$ $\rm{mag/arcsec}^2$ over a region with a 1.15 arcsec radius
 (\cite{mad90}).  For the
 CfA survey, we selected galaxies with apparent magnitudes brighter than
15.5 $B_J$ mag
 and whose surface brightness is at least 23.5 $B_J$
 $\textrm{mag/arcsec}^2$ over a region with a 4.5 arcsec radius--
criteria we consider only to be a ``reasonable'' estimate.  To mimic the
scatter in the relationship
 between Zwicky magnitudes and $B_J$ mag, we convolved the derived LF with
 a Gaussian of standard deviation 0.35 mag (\cite{bot90}).  For the
Las Campanas redshift
 survey (LCRS), we used the 112-fiber selection criteria given in Lin et
al.\ (1996)
 and took their magnitudes to be
 isophotal down to a surface brightness of 23 Gunn $r$ $\rm{mag/arcsec}^{2}$.
 For the Anglo-Australian Redshift Survey (AARS), we selected galaxies whose
apparent magnitude is
 less than 17.2 $B_J$ mag and whose surface brightness is at least
 23.5 $B_J$ $\rm{mag/arcsec}^{2}$.

 Using the $\chi^2$ test and taking $\sigma$ equal to
 $\sigma_{obs}(\sqrt{{N_{model}/N_{obs}}})$,  \footnote{Note that the
 observational mean is only
 an estimate of the
 true variance, which is determined by the model.} we compared the LFs
 predicted for various
 interpretations of the GK model to the actual determinations.
 Formally, the calculated LF for the GK model 
 (to within 0.3 magnitude of an observation) and our models are inconsistent
 with the LF of Mobasher et al.\ (1993),  Lin et al.\ (1996), 
 Loveday et al.\ (1992), and Marzke et al.\ (1994), to 8 $\sigma$, 31
 $\sigma$, 4 $\sigma$, and 5 $\sigma$, respectively.  Since these
 discrepancies are arguably a result of uncertainties in
 both the calibrations of
 the observed apparent magnitudes and the normalization due to the limited
 volumes surveyed, we will base our comparisons
 on those normalizations (only for CfA and AARS) and calibrations which
 produced the best fits to the calculated LFs.
 For these best-fit parameters, the LFs of our models are still generally
 inconsistent with the measured LF (the LF from the LCRS is inconsistent to
 19 $\sigma$), a result essentially due to the fact that the knees of
 the LFs are extremely well defined.  For the faint ($M_r > -19.4$) end
 of the LF, however, we find that only the LF from the LCRS is still
 inconsistent.   Our cycling models are especially inconsistent (Model A is
 inconsistent to 6 $\sigma$) as they predict
 many more faint galaxies than are observed (Table \ref{tbl-1}).

 We show the redshift distributions we predict for the Canada-France-Hawaii
 Redshift
 Survey (\cite{lil95}: CFHRS) for both the GK model and our models in Figure
 \ref{fig2}.
  In accordance with the selection procedure, we took the seeing FWHM to be
 0.9 arcsec and have included those galaxies
 which had a central surface brightness of  24.02 $I$ $\rm{mag/arcsec}^2$.
 We took their magnitudes to be isophotal down to a surface brightness
 of 27.52 $I$ $\rm{mag/arcsec}^2.$  
 The
 predicted overabundance at low redshifts has two sources:
 the very steep upturn at the faint end of the GK LF--a
 feature
 inherent to the GK model--and
 the additional populations of fading or cycling galaxies which are
 not forming stars.  Since this first source of low redshift galaxies
 predominates at $z<0.05$ while the second is more uniformly
spread over low redshifts, we have decided to examine the relative number of
galaxies observed and predicted for the redshift bins ($0<z<0.05$) and
($0.05<z<0.15$) separately.  For the sake of comparison, we assume
 the observed redshift distribution, though mildly incomplete,
 is representative.   

The GK model predicts too many galaxies in the lowest redshift
 bin (3.5 $\sigma$) but roughly the right number in the other
low redshift bin.  In contrast, all our models predict too
 many galaxies
 in this other low redshift bin (2.1-3.8 $\sigma$ inconsistency).
We have summarized these results in Table \ref{tbl-1}
along with the relative consistency levels of these models
 to the faint end of the LCRS $r$-band LF.  In Table
 \ref{tbl-1}, we have also included the
cumulative degree to which the CFHRS and the faint end of
 the LCRS $r$-band LF rule out the 
various models considered in this Letter.  Of course, one should
interpret these results with some caution as our analysis makes the
questionable assumption that the galaxies in the CFHRS are
unclustered.

Ignoring surface brightness selection effects, we
 have calculated number counts in the $B_J$ and $K$ bands for the GK model
and our models.  We display these calculations in
Figure \ref{fig3} along with a comparison to a set of recent 
observations.\footnote{Note that
for the purposes of this figure, the error bars on
 the number counts from \cite{met91} and \cite{met95} are equal to the sum
 in quadrature of half the estimated completeness correction and
the Poissonian error times one and a half.}
 The GK model and our
interpretations of it agree reasonably well with the observations in the
$B_J$ band, though the predictions seem to be about 25\% too high and low on
a portion of the bright and faint ends in the $K$ band, respectively. 
Though some of this difference can be attributed to the use of
different versions of the Bruzual \& Charlot spectral evolution code,
much of this difference simply results from the mildly imperfect fit
to the number counts used in producing the GK model.

To determine the sensitivity of the present results to the surface brightness
properties of these non-physical blue galaxies, we repeated our
calculations, assuming a surface brightness lower than Freeman's law by
1.5 $\rm{mag/arcsec}^2$.  In accordance with
expectations, we calculated that fewer fading galaxies and fewer cycling
galaxies would be observed in both the LFs considered and the CFHRS.
 For these lower surface brightness galaxies, we find that both the GK model
 and our models can be made consistent with the
 faint end of the LFs considered.  Nevertheless, the GK
 model and our models are still inconsistent with the number of galaxies
 at low redshift to 3.2 $\sigma$ and 4.1 $\sigma$, respectively.
  Therefore, while lowering the surface brightness of the non-physical blue
 galaxies in the GK model permits a reconciliation with the faint end of the
 Las Campanas LF, it
 does not permit a reconciliation with the lack of low redshift galaxies in
 the CFHRS.  Of course, one could always suppose these non-physical
 blue galaxies have even lower surface brightnesses than we have considered,
i.e. greater than 23.1 $B_J$ $\textrm{mag/arcsec}^2$,
 but at some point, this lower
surface brightness would remove these galaxies from other observations as
 well, such as the color distributions these non-physical blue galaxies
 were originally employed to explain.
 In fact, the observed properties of the galaxies responsible for the
 deep counts excess  only require a
steepening of the low luminosity end of the LF in 
the distant universe, $z\gtrsim 0.5$ (\cite{tre94}). 
 Fading by expansion,
 due perhaps to a very substantial wind that drives mass loss, might
 be invoked to reconcile the high redshift data with the local observations,
 but we have not explored this possibility (cf \cite{bab92}).

\section{Conclusions}

 In this Letter, we have proposed various physical interpretations of the
non-physical  population of blue galaxies in the GK model and have
calculated how these interpretations would be manifested
 in various determinations of the redshift distribution, the 
 luminosity function, and the number counts.
  Firstly,
we find that
the GK model and all our models predict too many galaxies at
low redshift ($0 < z < 0.05$) in the CFHRS--an excess which is intrinsic
to the GK model itself.  Glazebrook et al.\ (1995) previously reported this
 low redshift excess with regard to a similar model (\cite{koo93}).
 For our models, we predict an additional low redshift
 population which exceeds the observed
 galaxies in CFHRS in the redshift range ($0.05 < z < 0.15$).
Secondly, for our cycling models,
 we predict that too many intrinsically faint galaxies would be
 observed in the LCRS LF.  
If one supposes this bluest class of galaxies has lower surface
brightnesses, i.e.
23.1 $B_J$ $\textrm{mag/arcsec}^2$, we no longer find an excess of
 galaxies at the faint end of the LCRS LF.  Nevertheless, there is still a
discrepancy between the number of galaxies observed
 in the lowest redshift bins ($0 < z < 0.15$) and
 the number predicted from the GK model and our models, respectively. 
Hence passive evolution using ``normal'' galaxies, essentially in the spirit
of the GK model and the modifications that we have advocated,
fails to account both for the deep counts and for the low redshift
counterparts of the distant galaxies.
 One needs to add either luminosity evolution,
in the form for example of dynamical fading or a top heavy IMF, or number
evolution, as  occurs in merging histories,
 or some combination of these effects.

Recently, in a model which is somewhat similar to the GK model,
Pozzetti, Bruzual, \& Zamorani (1996) have
 presented an alternate set of models which propose to explain
much of the current observational data (number counts, redshift
 distributions, and color distributions) with essentially passive
evolution.  One of the most notable 
improvements of this new model over the GK model
is the relative absence of galaxies at low redshifts ($z < 0.05$).
Despite this improvement, this new model invokes a population of
eternally young (0.1 Gyr) galaxies quite similar to those galaxies used in
the GK model.
Making these galaxies physical in the ways outlined here
would have similar observational effects to those we have calculated, though
our own calculations have shown that the corresponding low redshift
effects are not large enough to cause a problem with the
observations employed in this paper.
The basic reason for this difference is that the non-physical blue
galaxies in this model make up a much smaller
fraction of the galaxies seen at any apparent magnitude than the
non-physical blue galaxies do in the GK model.  Nevertheless, for the case
that galaxies have ``normal'' surface brightnesses, this model still 
predicts a 200\% excess in the number of galaxies found at the faint 
($M_r \approx -18.6$) end of LF derived from the LCRS.

 As a  final note, while we have examined how both the GK model and
 our physical interpretations would manifest
 themselves in various determinations of
 the redshift distribution, number counts,
 and luminosity function at zero redshift, we suspect that it
 may also be fruitful
 to consider other constraints on passive evolution models
 from various recent studies that 
 have shown, albeit with sparser data,  that the luminosity function
 steepens (\cite{ell96}) or brightens (Lilly et al.\ 1995) as
 one progresses back in redshift space.

\acknowledgements

We are grateful to Huan Lin, Ron Marzke, Jon Loveday, and Bahram Mobasher
for sending
us information and/or data with regard to their determined luminosity
functions.  We would
also like to thank John Huchra for providing us with useful information on
various surveys and Caryl Gronwall for many helpful discussions and data
related to the GK work.  Rychard J. Bouwens gratefully acknowledges support
 from
an NSF Graduate Fellowship. This research has also been supported in part by
grants from NASA.

\clearpage

\begin{figure}
\caption{Luminosity functions calculated for the GK model and our models
(thick solid = GK model, dotted = Model A (rapid cycling), short
dashed = Model B (cycling), dashed = Model C (slow cycling), thin solid =
Model D (pure fading)).
For comparison, the LFs of Loveday et al.\ (1992), Marzke et al.\ (1994),
 Mobasher et al.\ (1993), and Lin et al.\ (1996) are all plotted
on this figure using circles, triangles, diamonds, and squares,
respectively.  See text for more details on models.
\label{fig1}}
\end{figure}

\begin{figure}
\caption{Predicted CFH redshift distribution based on the GK model and
our models (thick solid = GK model, dotted = Model A
(rapid cycling), short
dashed = Model B (cycling), dashed = Model C (slow cycling), thin solid =
Model D (pure fading)) with a comparison to CFHRS (Lilly et al.\ 1995)
displayed here as a histogram.  See
text for more detail on models.
\label{fig2}}
\end{figure}

\begin{figure}
\caption{Number counts in the $B_J$ and $K$ bands based on the GK model
and our models (thick solid = GK model, dotted = Model A
(rapid cycling), short
dashed = Model B (cycling), dashed = Model C (slow cycling), thin solid =
Model D (pure fading)) with a comparison to the observations of 
 Metcalfe et al.\ (1991), Metcalfe et al.\ (1995), Moustakas et al.\ (1995),
 Djorgovski et al.\ (1995), and Gardner et al.\ (1993)
displayed as solid circles, solid squares, asterisks, hollow squares, and
hollow triangles, respectively.  See text for more detail on models.  
\label{fig3}}
\end{figure}

\clearpage

\begin{deluxetable}{ccccc}
\tablecolumns{5}
\tablewidth{0pc}
\tablecaption{The inconsistency level of the present models with respect to
various observations. \label{tbl-1}}
\tablehead{
\colhead{} & \colhead{LCRS} & \colhead{CFHRS} & \colhead{CFHRS} & \colhead{} \\
\colhead{Model} & \colhead{Faint Gal.} & \colhead{\# predicted}  &
\colhead{\# predicted} & \colhead{Cumulative} \\
 & \colhead{Excess\tablenotemark{a}} & \colhead{$0<z<0.05$} &
\colhead{$0.05<z<0.15$} & \colhead{Inconsistency} \\
 \colhead{} & \colhead{} & \colhead{3 observed} & \colhead{22 observed} &
\colhead{Level\tablenotemark{b}} }
\startdata
GK & 6\% (1.4 $\sigma$) & 17 (3.5 $\sigma$) & 24 (0.4 $\sigma$)
 & 3.3 $\sigma$ \\
A & 68\% (5.3 $\sigma$) & 20 (3.8 $\sigma$) & 35 (2.1 $\sigma$)
 & 6.5 $\sigma$ \\
B & 43\% (3.2 $\sigma$) & 19 (3.6 $\sigma$) & 41 (2.9 $\sigma$)
 & 5.2 $\sigma$ \\
C & 0\% (0.9 $\sigma$) & 19 (3.7 $\sigma$) & 46 (3.5 $\sigma$)
 & 4.8 $\sigma$ \\
D & 11\% (1.3 $\sigma$) & 21 (3.9 $\sigma$) & 48 (3.8 $\sigma$)
 & 5.2 $\sigma$ \\
\enddata
\tablenotetext{a}{Percent excess of predicted galaxies over those observed
at the faint end ($M_r > -19.4$) of the LCRS (Lin et al.\ 1996)}
\tablenotetext{b}{Cumulative inconsistency of the model predictions with the
three observables in this table}
\end{deluxetable}

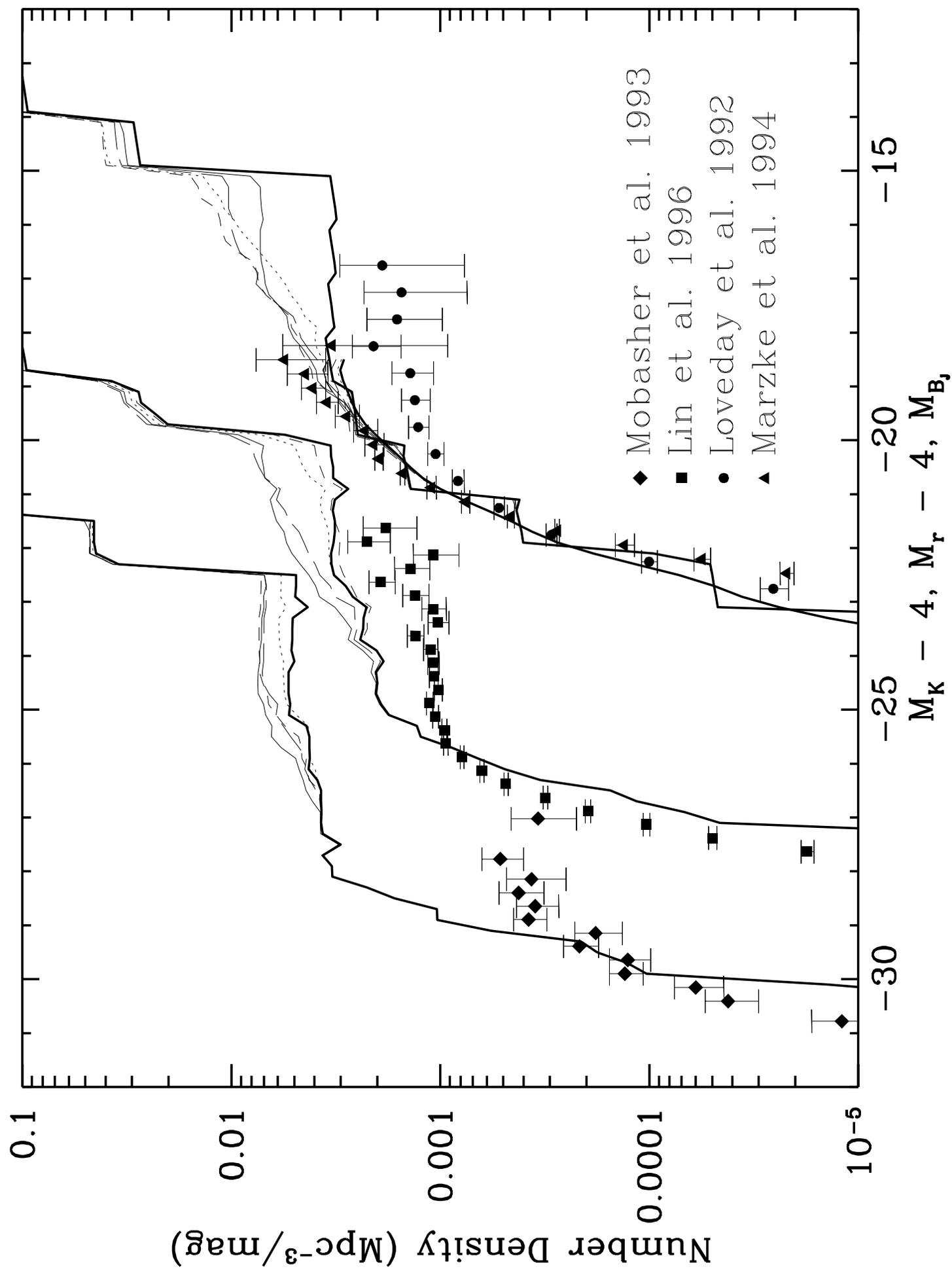
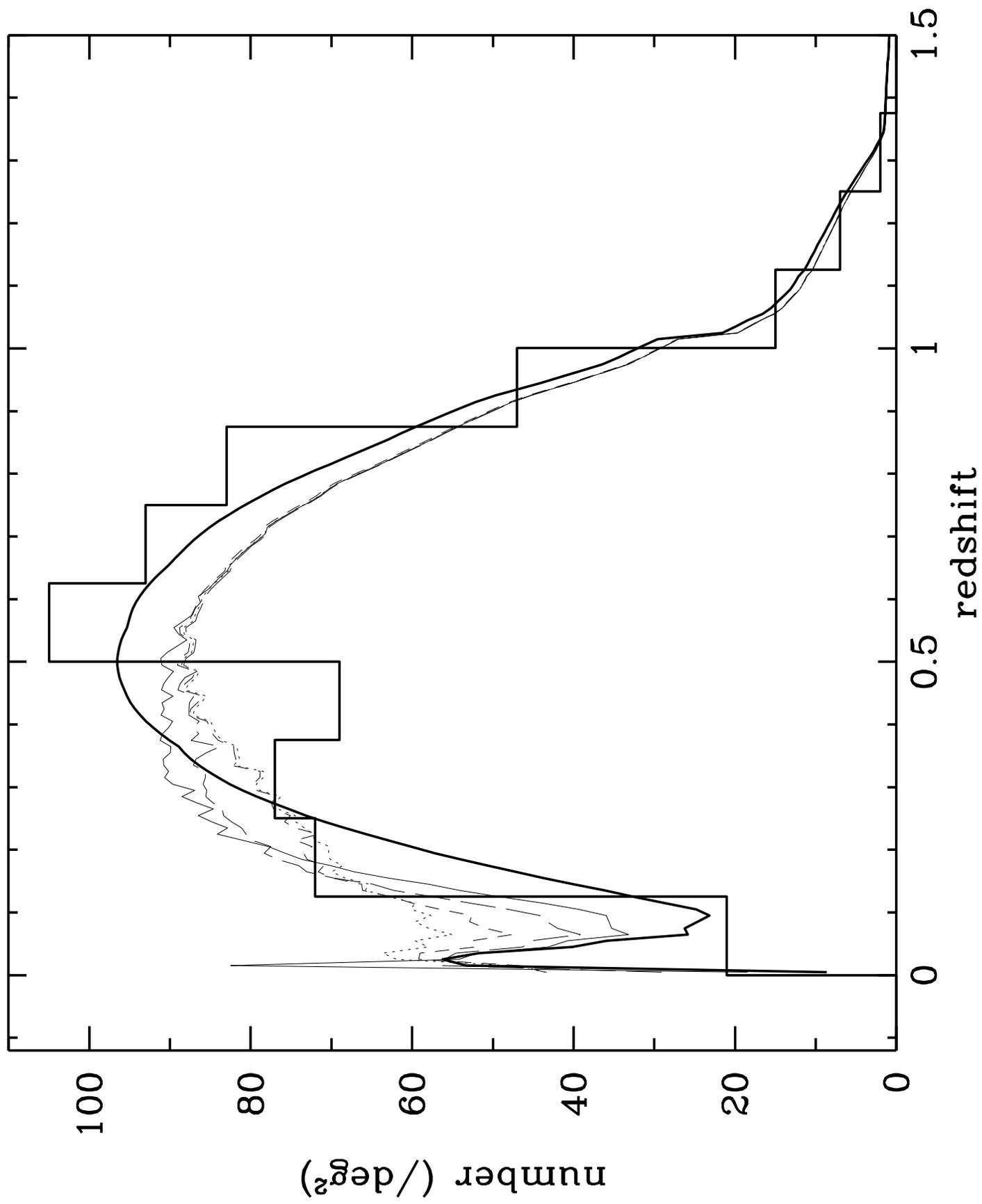
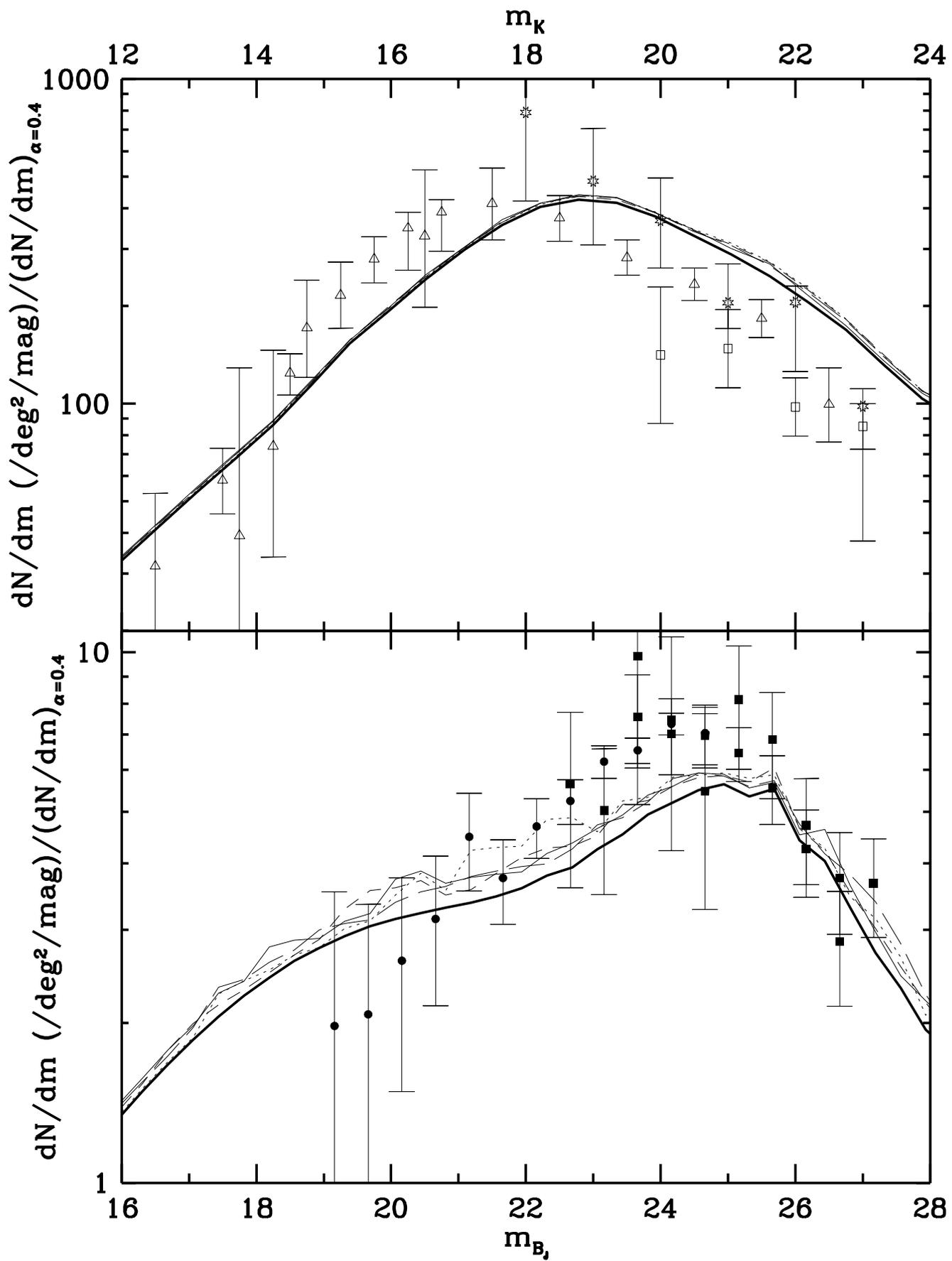
\end{document}